\documentclass[12pt,preprint]{aastex}

\slugcomment{Submitted to ApJ, 30 November 2007}

\shorttitle{Discreteness Effects in $\Lambda$CDM Simulations}

\shortauthors{A. B. Romeo et al.}

\begin{document}

\title{Discreteness Effects in Lambda Cold Dark Matter Simulations:\\
       A Wavelet-Statistical View}

\author{Alessandro B. Romeo}

\affil{Onsala Space Observatory,
       Chalmers University of Technology,
       SE-43992 Onsala, Sweden}

\email{romeo@chalmers.se}

\and

\author{Oscar Agertz, Ben Moore and Joachim Stadel}

\affil{Institute for Theoretical Physics,
       University of Z\"{u}rich,
       CH-8057 Z\"{u}rich, Switzerland}

\begin{abstract}

   The effects of particle discreteness in $N$-body simulations of
Lambda Cold Dark Matter ($\Lambda$CDM) are still an intensively
debated issue.  In this paper we explore such effects, taking into
account the scatter caused by the randomness of the initial
conditions, and focusing on the statistical properties of the
cosmological density field.  For this purpose, we run large sets of
$\Lambda$CDM simulations and analyse them using a large variety of
diagnostics, including new and powerful wavelet statistics.  Among
other facts, we point out (1) that dynamical evolution does not
propagate discreteness noise up from the small scales at which it is
introduced, and (2) that one should aim to satisfy the condition
$\epsilon\sim2d$, where $\epsilon$ is the force resolution and $d$ is
the interparticle distance.  We clarify what such a condition means,
and how to implement it in modern cosmological codes.

\end{abstract}


\section{INTRODUCTION}

   $N$-body simulations are becoming a more and more powerful tool for
investigating the formation of structure in the Universe.  Since this
is a grand-challenge problem with a huge dynamic range, it is of basic
importance to understand how significant numerical effects are in
cosmological simulations.  This is becoming even more important now,
with the advent of precision cosmology, which aims at breaking the
degeneracy of cosmological models and providing accurate determination
of the cosmological parameters at the percent level.  The difficult
task of simulations will then be to determine error bars in the output
data, taking into account not only the uncertainty of the measurements
themselves but also numerical effects, which enter the simulation
model and propagate during its evolution.

   Many recent investigations have focused on the effects of particle
discreteness in Lambda Cold Dark Matter ($\Lambda$CDM) simulations.
In fact, the number of particles $N$ determines two important
quantities: the mass resolution, and the wavelength of the
highest-frequency mode included in the initial conditions.  Such
quantities, together with the force resolution and other more
technical factors, affect the halo mass function and the statistical
properties of the cosmological density field (e.g., Splinter et al.\
1998; Knebe et al.\ 2000; Hamana et al.\ 2002; Smith et al.\ 2003;
Teodoro \& Warren 2004; Warren et al.\ 2006; Hansen et al.\ 2007;
Joyce \& Marcos 2007a; Baertschiger et al.\ 2008; Bagla \& Prasad
2008; Tinker et al.\ 2008), as well as the halo density profile and
the dynamics of structure formation (e.g., Knebe et al.\ 2000; Binney
\& Knebe 2002; Power et al.\ 2003; Binney 2004; Diemand et al.\ 2004;
Heitmann et al.\ 2005; Zhan 2006; Joyce \& Marcos 2007b; Vogelsberger
et al.\ 2008).  Wang \& White (2007) have shown that discreteness
effects are even more important in simulations of hot/warm dark
matter, where the initial power spectrum has a natural high-frequency
cut-off at a relatively long wavelength.  So why another contribution
concerning $\Lambda$CDM simulations?  Because the topic is still hot
(Melott 2007; Joyce et al.\ 2008; and references therein), and there
is more to learn:
\begin{itemize}
\item \emph{Actual significance of discreteness effects.}  As a
      cosmological simulation can only give one view of the local
      Universe, it is important to run an ensemble of simulations,
      varying the random realization of the initial conditions or, in
      other words, varying the phase and amplitude of the initial
      random fluctuations for a given power spectrum (Knebe \&
      Dom\'{\i}nguez 2003; Sirko 2005).  The randomness of the initial
      conditions causes statistical scatter in the diagnostics, which
      competes against the systematic effects of discreteness and
      should therefore be evaluated.  This requires running large
      ensembles of simulations.
\item \emph{Deeper and wider view of such effects.}  Wavelets are a
      state-of-the-art numerical technique used for extracting
      multiscale information from scientific data (see, e.g., Fang \&
      Thews 1998; Vidakovic 1999; Press et al.\ 2007, chap.\ 13.10).
      Despite their numerous applications in cosmology (e.g., Fang \&
      Feng 2000; He et al.\ 2005; Mart\'{\i}nez et al.\ 2005; Feng
      2007; Saar et al.\ 2007), they have not yet been used in the
      context of discreteness effects.  Wavelets provide a
      multiresolution view of the data, which in our case represent
      the density field.  The field is analysed first at the finest
      resolution consistent with the sampling, and then at coarser and
      coarser resolution levels.  Doing so, wavelets probe the
      structure of the field and the contributions from the various
      scales.  Besides, wavelets are sensitive to both the amplitude
      and the phase of the density fluctuations.  Thus wavelet-based
      statistics can offer a deeper and wider view of discreteness
      effects than traditional diagnostics.
\item \emph{Particular aspects of the problem.}  There are several
      aspects of the problem that deserve particular attention: (1)
      What is the range of scales affected by discreteness?  (2) As
      already mentioned, discreteness imposes two limitations: a
      finite mass resolution, and a lack of initial fluctuation power
      on small scales.  It would be interesting to study their
      statistical effects separately.  Binney (2004) himself concluded
      that ``it would be interesting to have a series of simulations
      in which the power spectrum is truncated at large wavenumbers,
      with the result that any increases in the particle number lead
      to the same structures being more and more densely populated,
      rather than to ever smaller-scale structures being simulated''.
      Although Moore et al.\ (1999) did not see a significant
      difference in the density profile of a halo simulated with a
      truncated power spectrum, Col\'{\i}n et al.\ (2008) found
      steeper central density profiles in a larger sample of haloes
      that formed with masses close to the free-streaming cutoff
      scale.  Here instead we study in detail how a truncated power
      spectrum affects the statistics of the density field.  (3)
      Discreteness effects also arise from the grid-like particle
      distribution used in the initial conditions (e.g., Hansen et
      al.\ 2007).  It would be worthwhile to check whether the initial
      grid leaves any statistical trace at low redshifts.  (4) A
      further aspect of the problem concerns the probability
      distribution of the initial fluctuations, and its evolution with
      redshift.
\end{itemize}

   For this purpose, we run two large sets of simulations.  In one we
vary $N$ and the random realization of the initial conditions, while
in the other we truncate the initial power spectrum and vary $N$
through different sampling techniques so as to further probe
discreteness effects and the transfer of power from large to small
scales.  In both sets we keep the force resolution fixed, so as to
decouple the effects of discreteness from those of force resolution
(cf.\ Binney 2004).  Our simulations span scales from
$80h^{-1}\mathrm{kpc}$ to $20h^{-1}\mathrm{Mpc}$.  We analyse the
statistical properties of the cosmological density field using a large
variety of diagnostics, including new and powerful wavelet statistics.
We compute all the diagnostics consistent with the force resolution,
so as to probe effects that are fully resolved dynamically.

   The rest of our paper is organized as follows.  The first set of
simulations is described in Sect.\ 2.  The actual significance of
discreteness effects against statistical scatter is assessed in Sect.\
3, where we also inquire into the nature of such scatter.  In Sect.\
4, we introduce the wavelet statistics and analyse discreteness
effects.  The second set of simulations is described in Sect.\ 5, and
particular aspects of the problem are probed in Sect.\ 6.  We draw the
conclusions in Sect.\ 7.

\section{SIMULATIONS}

   Our cosmological $N$-body simulations use the particle-mesh code by
Klypin \& Holtzman (1997), and are based on one of their runs.  We
have adopted the same code and basic run in an introductory study
(Agertz 2004).

   A particle-mesh code is appropriate for our purpose because it
computes all the dynamical quantities, from the density field
$\delta(\mbox{\boldmath $x$})=[\rho(\mbox{\boldmath
$x$})-\bar{\rho}]/\bar{\rho}$ to the forces, with given spatial
resolution.  The code by Klypin \& Holtzman (1997) is publicly
available and well described.  It can also be used for generating the
initial conditions and analysing the output data.  The initial
conditions are set up by using the Zeldovich approximation to displace
particles from a regular grid.  The power spectrum $P(k)$, correlation
function $\xi(r)$ and mass variance $\sigma_{M}^{2}(r)$ of the output
density field are computed consistent with the spatial resolution of
the code, which is twice the cell size $\Delta_{\mathrm{c}}$.

   The basic run is a $\Lambda$CDM simulation.  The cosmological
parameters are: $\Omega_{\Lambda}=0.7$, $\Omega_{\mathrm{m}}=0.3$,
$\Omega_{\mathrm{b}}=0.026$, $h=0.7$, $\sigma_{8}=1$ and $n=1$.  The
simulation has $N=32^{3}$ particles, $N_{\mathrm{c}}=128^{3}$ cells
and a box of $L=20h^{-1}\mathrm{Mpc}$.  It runs from redshift $z=15$
($a=0.0625$) to $z=0$ ($a=1$) in 469 steps ($\Delta a=0.002$).  This
simulation is simple and realistic enough for our purpose.

   The simulations of this paper have the same input parameters as the
basic run, except $N$ and the random number seed for generating the
initial conditions.  We use five values of $N$:
$N=16^{3},32^{3},\ldots,256^{3}$.  In other words, the number of
particles per cell ranges from $\frac{1}{512}$ to 8, and the spatial
resolution ranges from $\frac{1}{4}$ to 4 times the average
interparticle distance.  For each $N$, we generate ten random
realizations of the initial conditions.  Such a set of 50 simulations
is appropriate for exploring how their outcome depends on $N$, and for
evaluating the statistical scatter of the measurements.  Additional
simulations are discussed in Sect.\ 5.

\section{DISCRETENESS EFFECTS VS.\ STATISTICAL SCATTER}

\subsection{$P(k)$, $\xi(r)$ and $\sigma_{M}^{2}(r)$ Are
            Scatter-Dominated}

   Three popular statistical diagnostics used in cosmology are the
power spectrum $P(k)$, the correlation function $\xi(r)$ and the mass
variance $\sigma_{M}^{2}(r)$ (see, e.g., Peebles 1980; Coles \&
Lucchin 2002).  $P(k)$ is the average square amplitude of density
fluctuations on scale $2\pi/k$, with proper normalization.  In the
code by Klypin \& Holtzman (1997), $P(k)$ is computed as:
\begin{equation}
P(k)=\frac{1}{L^{3}}\,\frac{\sum|\hat{\delta}(\mbox{\boldmath $k$})|^
{2}}{\Delta N_{k}}\,,
\end{equation}
where $\hat{\delta}(\mbox{\boldmath $k$})$ is the fast Fourier
transform of the density field $\delta(\mbox{\boldmath $x$})$
tabulated in the mesh of the code, the sum is over all wavenumbers
spanning a spherical shell of radius $k$ and thickness $\Delta
k=2\pi/L$, $\Delta N_{k}$ is the number of harmonics in the shell, and
$L$ is the box size.  $\xi(r)$ is the real-space equivalent of $P(k)$
and measures the correlation strength of structures on scale $r$.  In
the code by Klypin \& Holtzman (1997), $\xi(r)$ is computed by
discretizing the relation:
\begin{equation}
\xi(r)=\frac{1}{2\pi^{2}}\int_{0}^{\infty}P(k)\frac{\sin(kr)}{kr}k^{2}
\mathrm{d}k\,.
\end{equation}
The algorithm consists of several steps, which cannot be translated
into a single formula [cf.\ routine FCORR(R) in PMpower.f].
Similarly, we compute $\sigma_{M}^{2}(r)$ by discretizing the
relation:
\begin{equation}
\sigma_{M}^{2}(r)=\frac{1}{2\pi^{2}}\int_{0}^{\infty}P(k)W^{2}(kr)k^{2
}\mathrm{d}k\,,
\end{equation}
where $W(kr)$ is the spherical top-hat window function:
\begin{equation}
W(x)=\frac{3}{x^{3}}(\sin x-x\cos x)\,.
\end{equation}
Note that the only difference between $\xi(r)$ and $\sigma_{M}^{2}(r)$
is the replacement of $\sin(kr)/kr$ with $W^2(kr)$.

   Another way to compute $\xi(r)$ and $\sigma_{M}^{2}(r)$ is by using
particle-based estimators (e.g., Knebe \& Dom\'{\i}nguez 2003), rather
than the mesh-based estimators above.  However, the computation takes
an order of magnitude longer time than the simulations themselves.
Besides, it is more difficult to compute $\xi(r)$ and
$\sigma_{M}^{2}(r)$ consistent with the spatial resolution of the
code.  This results in extra spikes, which are nothing but noise of
the particle-based estimation.  Apart from that, the results are
similar.  So in the following we go on discussing the mesh-based case.

   Another short digression.  In our simulations there are two Nyquist
frequencies involved.  One is the `particle Nyquist frequency'
\begin{equation}
k_{\mathrm{N}}=\pi N^{1/3}/L\,,
\end{equation}
which is associated with the grid-like particle distribution used in
the initial conditions.  This is the wavenumber at which the initial
power spectrum is truncated.  Therefore $k_{\mathrm{N}}$ determines
the initial number of modes.  The other Nyquist frequency, $\pi
N_{\mathrm{c}}^{1/3}/L$, is associated with the mesh of the code
($N_{\mathrm{c}}$ is the number of cells).  This is the largest
$|k_{i}|\ (i=1,2,3)$ that can be resolved in the mesh.  Harmonics with
$|k_{i}|>\pi N_{\mathrm{c}}^{1/3}/L$ are aliased into the principal
zone $|k_{i}|\leq\pi N_{\mathrm{c}}^{1/3}/L$, but they are greatly
attenuated if the mass-assignment scheme is TSC or CIC, as is used in
the code (see Hockney \& Eastwood 1988).  Among the two Nyquist
frequencies, only $k_{\mathrm{N}}$ varies in our set of simulations
and enters the following discussion.

   Let us then study how the number of particles affects the standard
diagnostics described above.  Fig.\ 1 shows $P(k)$, $\xi(r)$ and
$\sigma_{M}^{2}(r)$ in the range of scales spanned by the simulations,
that is approximately from the cell size
$\Delta_{\mathrm{c}}=L/N_{\mathrm{c}}^{1/3}$ to the box size $L$.  At
$z=15$, all the diagnostics manifest the peculiarity of the initial
conditions for $N<N_{\mathrm{c}}$.  Such discreteness effects appear
if initially the particles are distributed over a grid coarser than
the dynamical mesh.  At $z=0$, discreteness effects are hardly
detectable.  All the diagnostics show large statistical scatter
instead.  The initial and final redshifts are discussed in detail
below.
\begin{itemize}
\item \emph{Redshift $z=15$.}  For $N<N_{\mathrm{c}}$, we observe
      fluctuations in $P(k)$ beyond the particle Nyquist frequency
      $k_{\mathrm{N}}$.  As $k_{\mathrm{N}}$ increases with $N$, when
      there are one or more particles per cell we can no longer detect
      the peculiar imprint of the initial conditions on $P(k)$.
      Discreteness effects can also be observed in the other
      diagnostics for $N<N_{\mathrm{c}}$.  $\xi(r)$ fluctuates on
      scales smaller than a few times the average interparticle
      distance $L/N^{1/3}$.  $\sigma_{M}^{2}(r)$ changes slope on
      scales below $L/N^{1/3}$, and approaches the $r^{-4}$ behaviour
      expected for a grid-like distribution (e.g., Hansen et al.\
      2007).
\item \emph{Redshift $z=0$.}  The diagnostics are unaffected by the
      number of particles.  The only clear exception is $P(k)$ for
      $N\leq32^{3}$, which differs significantly from the other power
      spectra at wavenumbers larger than about five times its particle
      Nyquist frequency (its tail has a positive vertical offset with
      respect to the other tails).  The other power spectra are
      similar, considering their scatter, and so are all the
      correlation functions and mass variances.  On the other hand,
      the random realization of the initial conditions seems to have
      an important influence on all the diagnostics: their
      root-mean-square scatter is as large as a factor of two or more.
      In $P(k)$ and $\xi(r)$ the scatter is more important on large
      scales, while in $\sigma_{M}^{2}(r)$ it is quite uniform.  This
      difference between $\sigma_{M}^{2}(r)$ and $\xi(r)$ is a
      consequence of the large-scale behaviour of $W^{2}(kr)$, which
      decays faster than $\sin(kr)/kr$ and hence damps large-scale
      scatter more effectively.  Our findings reinforce those by Knebe
      \& Dom\'{\i}nguez (2003) and Sirko (2005), who observed large
      scatter in the standard diagnostics for a typical number of
      particles.
\end{itemize}

   The results of this analysis can be summarized as follows.  At low
redshifts the power spectrum, correlation function and mass variance
are dominated by statistical scatter, rather than by the systematic
effects of discreteness.  Discreteness effects are only detectable in
$P(k)$, far beyond the particle Nyquist frequency
($k\ga5\,k_{\mathrm{N}}$) and if there is much less than one particle
per cell ($N\leq32^{3}$ for $N_{\mathrm{c}}=128^{3}$).  In the other
diagnostics there is no clear dependence on the number of particles,
even if $N$ varies by three and a half orders of magnitude.  This is
in contrast to the clear dependence on $N$ and the small scatter at
high redshifts.

\subsection{$\delta(\mbox{\boldmath $x$})$ Is Discreteness-Dominated}

   Fig.\ 2 shows the density field $\delta(\mbox{\boldmath $x$})$
tabulated in the mesh of the code.  So the spatial resolution is the
same as for $P(k)$, $\xi(r)$ and $\sigma_{M}^{2}(r)$.  Here the effect
of varying $N$ is mainly a change in granularity:
$\delta(\mbox{\boldmath $x$})$ becomes much less granular as $N$
increases from $16^{3}$ to $64^{3}$ or $128^{3}$.  In contrast, the
random realization of the initial conditions turns out to influence
only the spatial pattern.%
\footnote{In Fig.\ 2, the random number seed for generating the
          initial conditions does not vary.  The density field shows
          different patterns because, as a consequence of increasing
          $N$, the realization of the fluctuations is different.  In
          fact, the code by Klypin \& Holtzman (1997) does not keep
          the same realization of long waves while adding more and
          more short waves.  But this phase difference does not change
          the granularity of the density field.}
Thus the results of Sect.\ 3.1 are at odds with the visual outcome of
the simulations.  At low as well as at high redshifts the density
field is dominated by the systematic effects of discreteness.  It is
peculiar, if not surprising, that standard statistics such as the
power spectrum, correlation function and mass variance are insensitive
to the granularity of the density field, even when this property
changes as much as in Fig.\ 2, whereas they are sensitive to its
random realization.

\subsection{The Scatter Can Be Regarded as Spurious and Be Reduced}

   In order to gain insight, we consider simpler statistics of the
density field, namely its standard deviation $\sigma$, skewness $S$
and kurtosis $K$ (the mean is zero).  $S$ and $K$ are useful for
measuring departures from Gaussianity (see, e.g., Press et al.\ 2007),
which are significant at low redshifts.  Using Numerical Recipes, we
compute $\sigma$, $S$ and $K$ from the density field
$\delta(\mbox{\boldmath $x$})$ tabulated in the mesh of the code.  The
analysis of these statistics shows that $\sigma$, $S$ and $K$ are also
scatter-dominated (see Fig.\ 3).

   Where does the degeneracy come from?  Fig.\ 4 illustrates what the
histogram of the density field looks like at redshift $z=0$.  In
particular, the top panel shows the histogram for the range of values
spanned by the density contrast, while the bottom panel shows a zoom
of the histogram (see figure caption).%
\footnote{We do not scale the $y$-axis logarithmically because it is
          the probability distribution, and not its logarithm, what
          enters into the definition of the statistics.}
The simplicity of this figure points out two remarkable peculiarities
of the $\delta$ distribution: (1) a huge spike at $\delta=-1$, and (2)
an extremely long tail for $\delta>0$.  In computing the statistics,
the tail of high density peaks has much more weight than the frequent
deep under-densities.  This makes the statistics very sensitive to
rare events, and hence scatter-dominated.  But how genuine is such
scatter?  We know that it is difficult to extract robust statistical
information from data characterized by a long-tailed probability
distribution (see, e.g., Press et al.\ 2007).  Robust estimation
requires defining appropriate statistics (see Press et al.\ 2007), or
even transforming the data (see Stuart \& Ord 1991).  Thus the
sensitivity pointed out above means that standard statistics of the
density field are not so well defined, and that their scatter can be
regarded as spurious.

   Considering deep under-densities and high density peaks separately,
in view of their distinct cosmological meaning (voids and dark matter
haloes), still yields a strongly unbalanced $\delta$ distribution:
clustered below $\delta\sim0$ and dispersing up to $\delta\sim200$.
What about a statistical transformation of the density field?  Let us
consider the $\ln(1+\delta)$ distribution, for $\delta\neq-1$.  This
transformation shortens the over-density tail, and it even makes the
distribution roughly normal (cf.\ Fig.\ 5).  In addition, this
transformation is appealing because $1+\delta$ is a basic cosmological
quantity and the $\ln(1+\delta)$ distribution matches the $\delta$
distribution in the linear regime.  Lognormal models have been
discussed by Coles \& Jones (1991) and Kayo et al.\ (2001) among
others.  One may find better transformations considering the class of
$\frac{1}{\alpha}\,[(1+\delta)^{\alpha}-1]$ distributions, for
$\delta\neq-1$ and $0<\alpha<\frac{1}{2}$, but at the cost of a free
parameter to fine-tune.  We do not follow that approach.

   Encapsulating the singularity of the $\delta$ distribution
($\delta=-1$) in one statistic, the fraction of void cells or void
probability $\nu$, we can define statistics of the transformed density
field such as its mean $\mu$, standard deviation $\sigma$, skewness
$S$ and kurtosis $K$ (we reuse old symbols for denoting the new
quantities).  Note that $\nu$ is not meant to describe the
distribution of voids as cosmological structures, which would require
a topological approach.  Note also that now $\mu\neq0$, since the
transformation has a bias (see Stuart \& Ord 1991), and $S$ and $K$
measure departures from lognormality.  Such statistics are
discreteness-dominated, as is the density field itself: a sharp trend
with $N$ emerges and the scatter is mostly unnoticeable, at low as
well as at high redshifts (cf.\ Fig.\ 6).  In particular, the fact
that the fraction of void cells decreases with the number of particles
for a given cell size is in natural agreement with the visual outcome
of the runs.

   Summarizing, in this section we have learned that the statistical
scatter can be regarded as spurious and be strongly reduced.  The
consideration of scale-dependent statistics adds complexity but does
not change the message.  In Sect.\ 4, we will see that the method of
scatter reduction can be extended to wavelet statistics, which is
fundamental for an appropriate multiscale analysis.

\section{WAVELET-STATISTICAL ANALYSIS}

\subsection{The Fast Wavelet Transform}

   Data such as the cosmological density field enclose information on
various scales.  In order to extract such information, we should be
able to separate small-scale features from large-scale features and to
understand their contributions to the overall structure of the data.
In this section we describe a technique that can be used for the
purpose above: the fast wavelet transform.  For further reading see
Romeo et al.\ (2003), hereafter Paper I, and Romeo et al.\ (2004),
hereafter Paper II.  In particular, Paper II provides a
reader-friendly and self-contained discussion of wavelets, from the
basics to advanced aspects of the technique.

   The fast wavelet transform involves localized wave-like functions,
which are dilated over the relevant range of scales and translated
across the data.  The contributions of small-scale and large-scale
features are singled out with an iterative procedure.  The first step
consists of separating the smallest-scale features from the others.
It is done by passing the data through a high-pass filter and a
complementary low-pass filter.  These filters are the discrete
counterparts of the analysing functions of the transform, the wavelet
$\psi(x)$ and the scaling function $\phi(x)$, respectively, and are
constructed with a mathematical technique known as multiresolution
analysis.  Filtering produces redundant information, since each set of
filtered data has the same size as the original data.  Redundancy is
avoided by rejecting every other point of the filtered data.  It is
well known that down-sampling produces aliasing in the context of the
Fourier transform, but the filters of the wavelet transform are
constructed in such a way as to eliminate it.  The second step
consists of separating the features that appear on a scale twice as
large as in the first step.  It is done by regarding the low-pass
filtered and down-sampled data as new input data, and by analysing
them as in the first step.  The procedure continues until all features
below a given `upper scale' are separated.  In summary, the 1D fast
wavelet transform decomposes the original data into a coarse
`approximation' and a sequence of finer and finer `details', keeping
the total size of the data constant (cf.\ Paper I, fig.\ 2).  The 2D
or 3D fast wavelet transform is similar to the 1D case, except for the
more complicated structure of the transformed data (cf.\ Paper II,
fig.\ 6).  In general, given $n$D data of size $N_{\mathrm{d}}^{n}$,
the first step of the transform decomposes them into $2^{n}$ parts of
size $(N_{\mathrm{d}}/2)^{n}$: 1 approximation and $2^{n}-1$ details,
one for each axis and each diagonal.  This is done by 1D transforming
the data along each index, for all values of the other indices,
consecutively.  The second step decomposes the approximation into
$2^{n}$ parts of size $(N_{\mathrm{d}}/4)^{n}$, and so on.

   \emph{Note} that there is an important difference between the
approximation and each of the details produced by the fast wavelet
transform.  Independent of the number of dimensions, each detail is a
compact piece of information concerning a single scale.  In contrast,
the approximation encloses (unprocessed) information on various
scales; it can be viewed as a smoothed miniature of the data.  Note,
however, that there is a non-standard version of the multidimensional
discrete wavelet transform where the details have mixed scale content.
That is the one described in Numerical Recipes [see in particular
fig.\ 13.10.4\,(b) of the third edition (2007)].

\subsection{Transforming the Density Field and Computing Its
            Statistics}

   We compute the fast wavelet transform of the output density field
in each simulation by using the Code JOFILUREN (Papers I and II), and
refer to Paper II for a thorough discussion of the method.  To do the
computation, we must specify the analysing function and the upper
scale of the fast wavelet transform.  We choose the `bior\,4.4'
wavelet (see Paper II, fig.\ 1), which is the one suggested in Paper
II for cold dark matter simulations.  The upper scale is specified in
terms of the scale parameter $N_{\mathrm{t\,min}}$: an upper scale of
$2^{n}$ cell sizes corresponds to
$N_{\mathrm{t\,min}}=N_{\mathrm{c}}^{1/3}/2^{n}$, where
$N_{\mathrm{c}}$ is the number of cells.  We set
$N_{\mathrm{t\,min}}=16$, which is the lower bound suggested in Paper
II for the `bior\,4.4' wavelet.  Hence the density field is
wavelet-analysed at spatial scales $2^{s-1}\Delta_{\mathrm{c}}\
(s=1,\ldots,4)$, where $\Delta_{\mathrm{c}}$ is the cell size.  The
corresponding details $\mathcal{D}_{s}$ are sets of
$7N_{\mathrm{c}}/8^{s}$ data $D_{s}(i,j,k)$,%
\footnote{We are neglecting the 3D substructure of the details, here
          irrelevant.  Hence, at a given $s$, the spatial indices
          $(i,j,k)$ span the cube $\mathcal{C}_{s-1}=\{1\leq i,j,k\leq
          N_{\mathrm{c}}^{1/3}/2^{s-1}\}$ minus its subset
          $\mathcal{C}_{s}$.}
which can be used for probing the statistical properties of the
density field at such spatial scales.  The approximation is less
useful for this purpose because of its mixed scale content, as noted
in Sect.\ 4.1.

   Wavelet statistics of the density field should not be computed
directly from $\mathcal{D}_{s}$.  In fact, the distribution of $D_{s}$
values at redshift $z=0$ shows a central singularity and a very long
tail on both sides.  Such features are similar to those discussed in
Sect.\ 3.3, and have similar consequences.  Therefore we extend the
procedure followed in that case.  Using $\ln[1+\delta(\mbox{\boldmath
$x$})]$, rather than $\delta(\mbox{\boldmath $x$})$, does not solve
the problem here because this field diverges in a significant fraction
of the mesh (the void cells) and hence its fast wavelet transform is
not defined.  Our solution is to consider the subsets
$\mathcal{D}_{s}^{0}=\{D_{s}(i,j,k)=0\}$ and $\{D_{s}(i,j,k)\neq0\}$
separately, and transform the latter:
$\mathcal{D}_{s}^{\mathrm{T}}=\{\ln|D_{s}(i,j,k)|\}$.  We then define
the void probability $\nu_{s}$ as the fraction of vanishing detail
coefficients at spatial scale $s$.  We compute $\nu_{s}$ by counting
the number of data contained in $\mathcal{D}_{s}^{0}$, and recalling
that there are $7N_{\mathrm{c}}/8^{s}$ detail coefficients in total
(at scale $s$).  We also compute the mean $\mu_{s}$, standard
deviation $\sigma_{s}$, skewness $S_{s}$ and kurtosis $K_{s}$ of the
data contained in $\mathcal{D}_{s}^{\mathrm{T}}$.  Such wavelet
statistics are a scale-dependent generalization of the statistics
discussed in Sect.\ 3.3.  A generic member of this family is denoted
by $f_{s}$.  Hereafter, the subscript `$s$' is added only when needed.

\subsection{A First View of Discreteness Effects}

   Fig.\ 7 shows the wavelet statistics as functions of the number of
particles at various spatial scales.  Note at once that the large
scatter of $S_{4}(N)$ and $K_{4}(N)$ is not a failure of our
statistical transformation.  It appears because the skewness and the
kurtosis are high-order statistics, and at that spatial scale there
are relatively few detail coefficients.  By analysing the behaviour of
$f_{s}(N)$ at $z=15$, we learn that the imprint of the initial
conditions on the wavelet statistics is twofold.  First, there is a
strong correlation between the number of particles and the spatial
scale: $f_{s}(N\,8^{n})$ behaves approximately as $f_{s+n}(N)$.  In
fact, the red curves shifted by one step to the left match the green
curves, which in turn match the blue ones, and so on.  This
correlation appears because the effect of varying $N$ is, to
zeroth-order approximation, a simple change of scale in the particle
distribution.  Second, at a given spatial scale $s\leq3$, there is a
number of particles $N=64^{3}/8^{s-1}$ that minimizes $\sigma$ and
$S$, and maximizes $K$.  This follows from the fact that the $\delta$
distribution has a transition for $N=64^{3}$ (see Sect.\ 6.2).  The
behaviour of $f_{s}(N)$ at $z=0$ is more difficult to understand in
detail.  Nevertheless, we can deduce two basic facts:
\begin{enumerate}
\item There is a weak trace of the initial $N$--$s$ correlation, while
      there is no critical $N$.  This suggests that the average
      interparticle distance is still a significant scale when the
      particle distribution is hierarchically clustered.
\item Independent of redshift, the behaviour of $f_{s}(N)$ simplifies
      at $s\geq3$: all the wavelet statistics converge for large $N$,
      and are possibly constant at $s>4$.  In other words, increasing
      $N$ affects smaller and smaller spatial scales; scales larger
      than $s=4$ (the average interparticle distance for $N=16^{3}$)
      are possibly unaffected.  This suggests that discreteness
      effects are confined to scales smaller than about twice the
      average interparticle distance, and do not propagate bottom-up
      while cosmological structures form.  If confirmed, this is one
      important aspect of the robustness of cosmological $N$-body
      simulations.
\end{enumerate}

\section{ADDITIONAL SIMULATIONS}

   To understand more, we carry out three additional sets of
simulations, which are intermediate between those with $N=16^{3}$ and
$N=256^{3}$.  Each set is a statistical ensemble of ten simulations
produced by varying the random realization of the initial conditions,
as before.  The number of particles and the other input parameters are
the same as in the original $N=256^{3}$ simulations.  What differs in
each set is discussed below.

   In the first set, the initial power spectrum $P_{\mathrm{i}}(k)$ is
truncated at wavenumber $k=k_{\mathrm{max}}=16\pi/L$, where $L$ is the
box size.  One simulation is shown in Fig.\ 8 (top).  Recall that the
natural truncation wavenumber is the particle Nyquist frequency
$k_{\mathrm{N}}=\pi N^{1/3}/L$, which determines the initial number of
modes $N_{\mathrm{m}}\approx(k_{\mathrm{N}}L/\pi)^{3}\approx N$.
Hence $k_{\mathrm{max}}$ can be regarded as an effective particle
Nyquist frequency that sets the initial number of modes to
$N_{\mathrm{m}}\approx(k_{\mathrm{max}}L/\pi)^{3}\approx16^{3}$, while
the number of particles is $N=256^{3}$.  Comparing this set with the
$N=16^{3}$ and $N=256^{3}$ simulations, we can then study the
behaviour of the statistics as the mass resolution and the initial
number of modes vary independently.

   In the second set, the output density field is computed by
sub-sampling the particle distribution regularly, selecting $16^{3}$
particles (loop over particle index and choose every 16th particle).
One simulation is shown in Fig.\ 8 (middle).  Sub-sampling means loss
of information at scales smaller than about twice the new average
interparticle distance (sampling theorem).  In addition, regular
sub-sampling leaves the initial particle distribution grid-like.  So,
if this set turns out to be similar to the $N=16^{3}$ simulations, it
means that discreteness effects imply a loss of information similar to
sub-sampling, and hence that their spatial range is about twice the
average interparticle distance.

   In the third set, the output density field is computed by
sub-sampling the particle distribution randomly, selecting $16^{3}$
particles (see the selection sampling technique by Knuth 1998).  One
simulation is shown in Fig.\ 8 (bottom).  Random sub-sampling makes
the initial particle distribution Poisson-like.  Comparing this set
with the regularly sub-sampled simulations, we can then check whether
the initial grid leaves any statistical trace at low redshifts.

\section{PROBING DISCRETENESS EFFECTS}

\subsection{Spatial Range and Complexity}

   Fig.\ 9 shows the wavelet statistics as functions of the spatial
scale for five sets of simulations: the original $N=16^{3}$ and
$N=256^{3}$ simulations; and the additional simulations with
power-spectrum truncation, regular and random sub-sampling.  Recall
that $N=16^{3}$ / $N=256^{3}$ represents the case in which the spatial
resolution scale ($s=2$) is much smaller/larger than the average
interparticle distance ($s=4$ / $s=0$).  Our deductions are the
following:
\begin{enumerate}
\item The $N=256^{3}$ simulations are more basic than the others.  The
      wavelet statistics have featureless behaviour at high redshifts.
      The behaviour at low redshifts is also featureless, except that
      the standard deviation and the kurtosis depart from monotonicity
      below the spatial resolution scale.  Another interesting aspect
      of the evolution with redshift is that the skewness and the
      kurtosis approach zero, which means that the density field
      becomes approximately lognormal.
\item A comparison between the simulations with power-spectrum
      truncation and the $N=256^{3}$ simulations illustrates how
      complexity arises if the effective particle Nyquist scale is
      spatially resolved ($s=5$): at high redshifts the standard
      deviation, the skewness and the kurtosis oscillate; at low
      redshifts they all depart from monotonicity below the spatial
      resolution scale.  Decreasing the number of modes also yields a
      systematic decrease or increase in the wavelet statistics, with
      one exception.  A further interesting aspect of the evolution
      with redshift is that the simulations with power-spectrum
      truncation become more similar to the $N=256^{3}$ simulations.
      This means that there is transfer of statistical information
      from the modes initially excited to those initially suppressed,
      but the loss of information is still significant at $z=0$.  If
      not only the effective particle Nyquist scale but also the
      average interparticle distance is spatially resolved, then
      further complexity arises ($N=16^{3}$ simulations vs.\
      simulations with power-spectrum truncation).  The peculiarities
      of the wavelet statistics are pointed out in Sect.\ 4.3.
\item The regularly sub-sampled simulations agree rather well with the
      $N=16^{3}$ simulations.  This confirms that discreteness effects
      are insignificant beyond a scale of about twice the average
      interparticle distance.
\item At $z=15$, random sub-sampling differs significantly from
      regular sub-sampling.  It suppresses the minimum of the standard
      deviation and of the skewness, and the maximum of the kurtosis
      and of the mean at a scale of half the effective average
      interparticle distance ($s=3$).  It also increases the void
      probability at the cell-size scale by more than a factor of
      three.  At $z=0$, random sub-sampling is equivalent to regular
      sub-sampling.  Therefore the initial grid has a strong imprint
      on the wavelet statistics at high redshifts, whereas there is no
      memory of the initial grid at low redshifts.
\end{enumerate}

\subsection{Initial Non-Gaussianity from Gaussian Initial Conditions}

   In cosmological $N$-body simulations, the initial conditions are
generated assuming that the random density field is Gaussian.
Gaussianity is one of the basic cosmological assumptions.  It implies
that the density field is entirely characterized by its power spectrum
or correlation function.  (See, e.g., Peacock 1999.)

   But is the density field resulting from the initial conditions
really Gaussian?  Fig.\ 10 illustrates what the histogram looks like
when we vary the number of particles.  Each histogram is shown for the
range of values spanned by the density contrast, and for the subrange
$-1\leq\delta\leq1$.  Apart from the peculiarities pointed out in
Sect.\ 3.3, note that the $\delta$ distribution has a transition for
$N=64^{3}$: it is one-sided for $N\leq64^{3}$ and two-sided for
$N>64^{3}$.  This means that, if the spatial resolution scale is
smaller than twice the average interparticle distance, as usual, then
the initial density field estimated by the code is markedly
non-Gaussian (though the field that the point particles are sampling
is Gaussian).  The same is true even when we start the simulations as
early as at $z=100$ (cf.\ Fig.\ 11).  Such inconsistency arises
because, at that high force resolution, the mass distribution looks
granular: there is an excess of both high density peaks and deep
under-densities.  Such initial non-Gaussianity is what this or other
codes actually `see' at that high force resolution; and its effects
will propagate dynamically.  At lower resolution, the departure from
Gaussianity is moderate at $z=15$ and small at $z=100$ (cf.\ Figs 10
and 11).  Technical issues are discussed in Sect.\ 7.

\section{CONCLUSIONS}

   The significance of discreteness effects in $\Lambda$CDM
simulations depends on two comoving spatial scales: the force
resolution $\epsilon$, and the average interparticle distance $d$.
Here $\epsilon$ is also the resolution scale for density and
statistical estimation.  Our wavelet analysis shows that discreteness
has a strong impact if $\epsilon\ll2d$:
\begin{itemize}
\item The simulations are inconsistent with one of the basic
      cosmological assumptions.  In fact, the initial random density
      field is markedly non-Gaussian, even though is assumed to be
      Gaussian in the initial conditions (Sect.\ 6.2).
\item At low redshifts the density field departs significantly from
      lognormality and further complexity arises (Sect.\ 6.1).
\item The comoving spatial scales $s$ affected by discreteness span
      the wide range $\epsilon\la s\la2d$ (Sect.\ 6.1; see also Sect.\
      4.3).
\end{itemize}
Discreteness effects become insignificant if $\epsilon\sim2d$.  This
condition guarantees that the statistical properties of the
cosmological density field are modelled accurately throughout the
range of scales spanned by the simulation.  In particular, this is
fundamental for probing the imprints of primordial non-Gaussianities
on large-scale structure, which is a topic of current interest (e.g.,
Dalal et al.\ 2007; Grossi et al.\ 2008; Hikage et al.\ 2008).  These
results have two implications.  One is that $2d$, and not $\epsilon$,
is the minimum scale for extracting robust statistical information
from the simulation data.  The other concerns the trade-off between
force and mass resolution in modern cosmological codes, which is
discussed below.

   Let us consider Particle-Mesh (PM) codes using Adaptive Mesh
Refinement (AMR), such as ART (Kravtsov et al.\ 1997), MLAPM (Knebe et
al.\ 2001), RAMSES (Teyssier 2002), and others.  The condition
$\epsilon\sim2d$ can be implemented adaptively by requiring that there
are no less than $n$ particles per cell, where $n$ depends on the
mass-assignment/force-interpolation scheme.  In fact,
$\epsilon\sim\Delta_{\mathrm{c}}$ for NGP,
$\epsilon\sim2\Delta_{\mathrm{c}}$ for CIC and
$\epsilon\sim3\Delta_{\mathrm{c}}$ for TSC, where
$\Delta_{\mathrm{c}}$ is the cell size (for a description of these
schemes see Hockney \& Eastwood 1988).  In addition,
$n\sim(\Delta_{\mathrm{c}}/d)^{3}$.  Hence $n\sim8$ for NGP, $n\sim1$
for CIC and $n\sim\frac{1}{3}$ for TSC.  Such values are comparable to
those currently used.  This means that our condition is easy to
fulfil, and hence discreteness effects can be kept under control in
AMR codes.

   In tree-based codes such as GADGET-2 (Springel 2005) and PKDGRAV
(Stadel 2001), the force resolution is equal everywhere.  The
criterion for resolving the small-scale dynamics of structure
formation is then more demanding and imposes $\epsilon\ll2d$, as is
currently used.  This means that it may not be easy to enforce such a
criterion and our condition together, although their domains of
applicability are complementary.  An interesting possibility would be
to let the force resolution vary with redshift so as to enforce such
requirements in distinct regimes of clustering, following the transfer
of power from large to small scales.  A more radical alternative is to
have adaptive force resolution, as in the case of AMR codes.  A
variable softening length can now be implemented in a form that
conserves momentum and energy exactly (Price \& Monaghan 2007), and
its use has also been suggested in other contexts (e.g., Bate \&
Burkert 1997; Dehnen 2001; Nelson 2006; Price \& Monaghan 2007;
Wetzstein et al.\ 2008).

\acknowledgments

   We are very grateful to Anatoly Klypin for making his particle-mesh
code publicly available and for help; and to Elena D'Onghia, Michael
Joyce, Alexander Knebe, Lucio Mayer, Francesco Miniati, Andrew Nelson
and Volker Springel for valuable discussions.  We are also grateful to
an anonymous referee for constructive comments and suggestions, and
for encouraging future work on the discreteness topic.  The first
author thanks the wonderful hospitality and strong encouragement of
the Institute for Theoretical Physics at the University of Z\"{u}rich.
He also acknowledges the financial support of the Swedish Research
Council, and the ASTROSIM short-visit grant 1815 by the European
Science Foundation.

\clearpage

\begin{figure}
\epsscale{.84}
\plotone{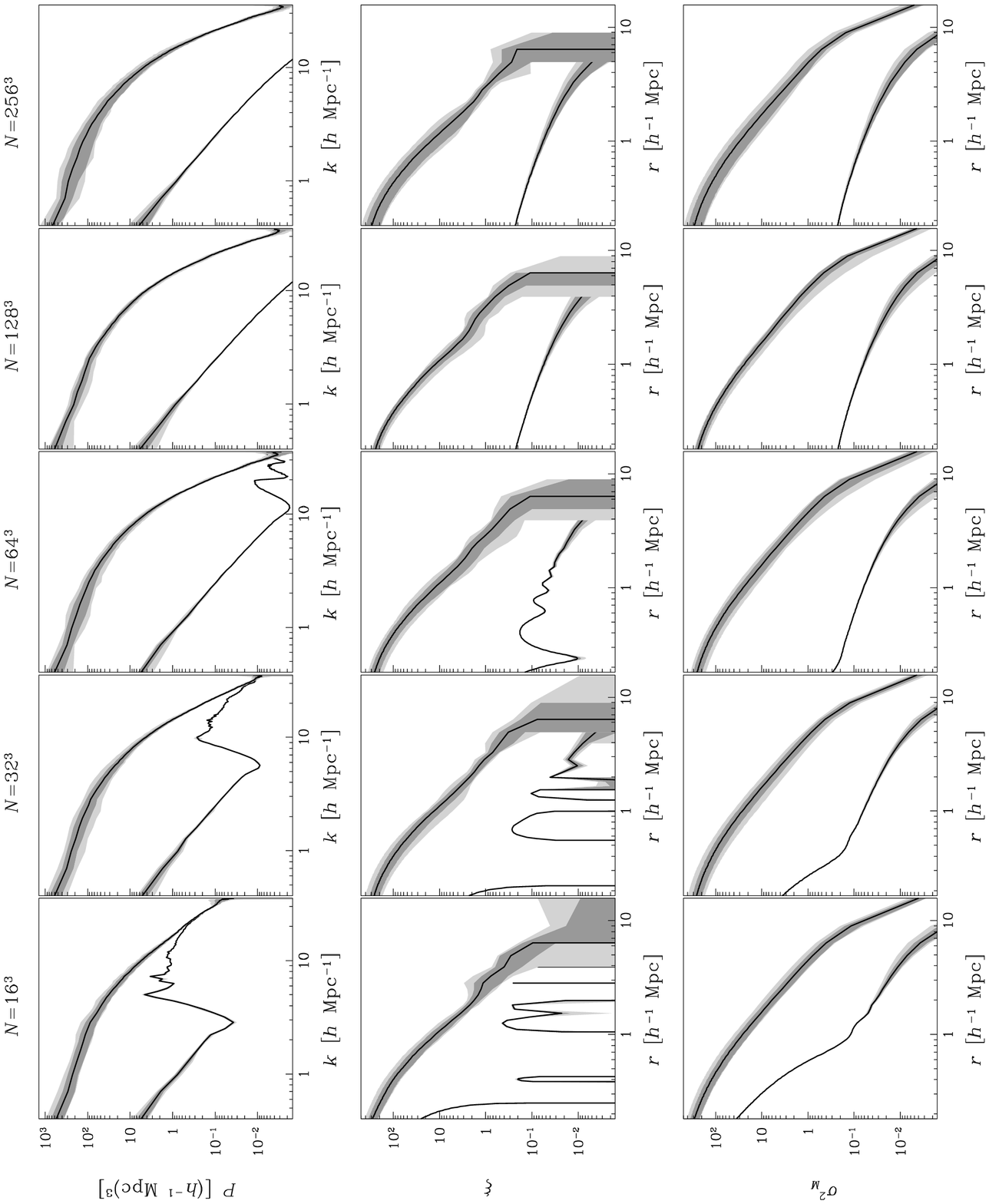}
\caption{The power spectrum $P(k)$, correlation function $\xi(r)$ and
         mass variance $\sigma_{M}^{2}(r)$ for various $N$.  Each
         panel shows the statistical scatter of the diagnostic at
         redshifts $z=15$ (lower set of curves) and $z=0$ (upper set
         of curves).  The black line, dark and light grey regions
         represent the average, root-mean-square deviation and range
         of the diagnostic, respectively.}
\end{figure}

\begin{figure}
\epsscale{1.}
\plotone{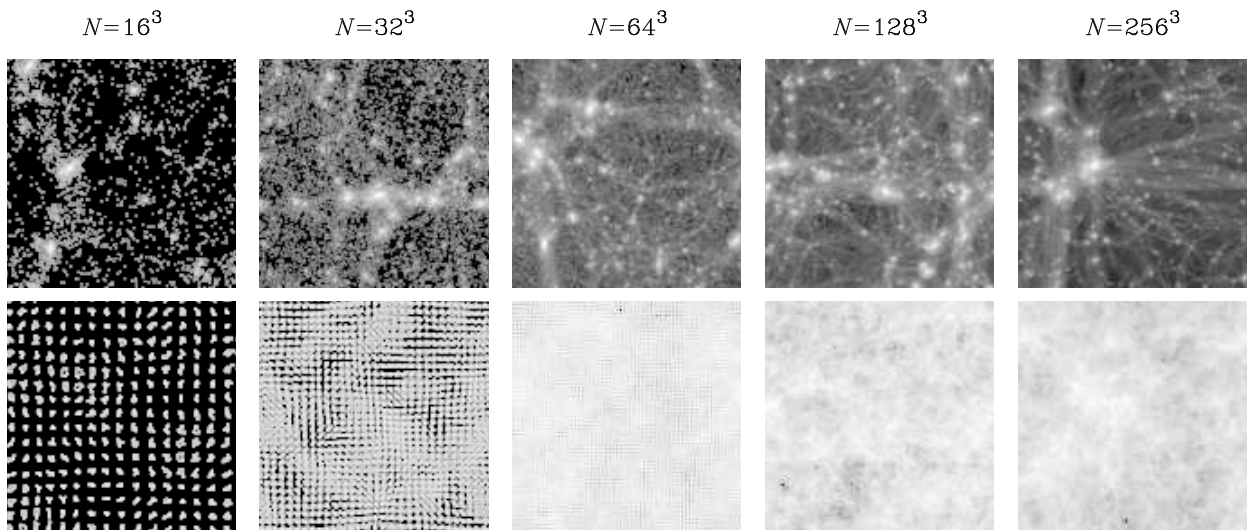}
\caption{Logarithmic grey-scale map of the projected density field for
         various $N$ at redshifts $z=0$ (top) and $z=15$ (bottom).}
\end{figure}

\begin{figure}
\epsscale{.54}
\plotone{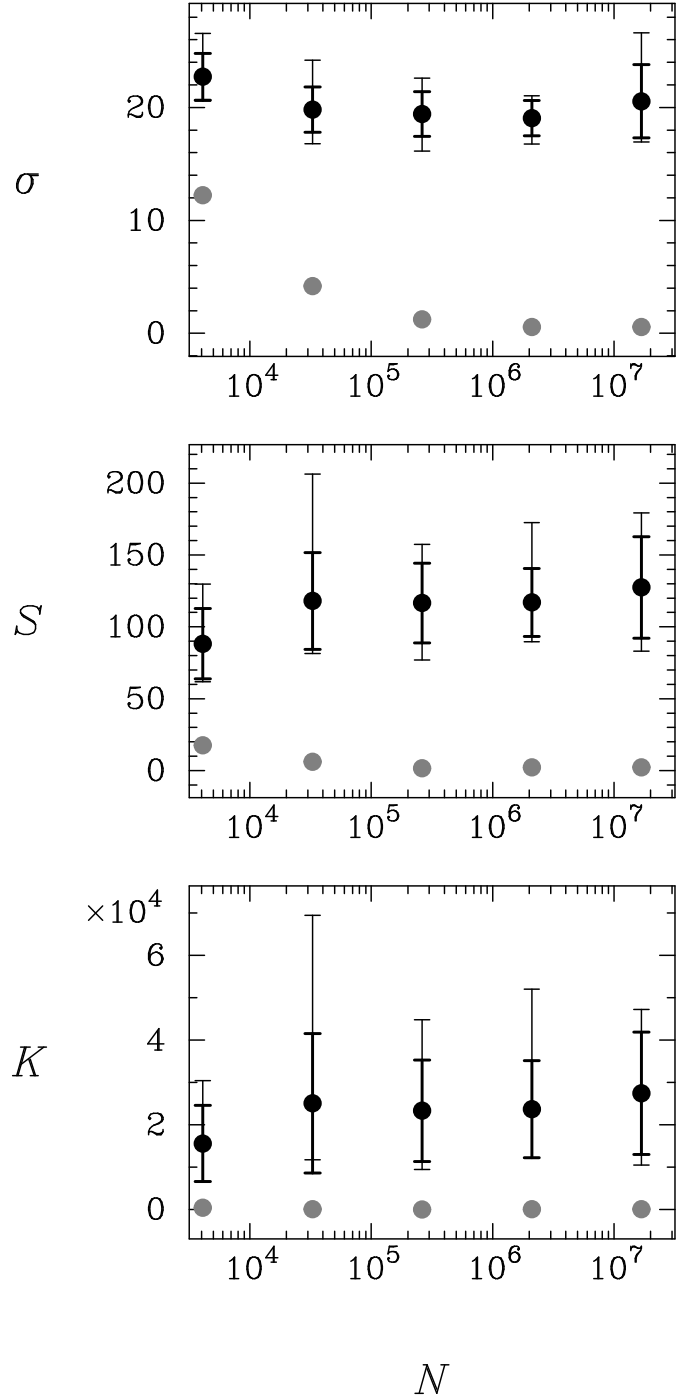}
\caption{Statistics of the density field as functions of $N$: the
         standard deviation $\sigma$, skewness $S$ and kurtosis $K$.
         The redshifts shown are $z=0$ (black) and $z=15$ (grey).  The
         scatter of the diagnostics is shown by plotting their average
         (circles), root-mean-square deviation (thick error bars) and
         range (thin error bars).}
\end{figure}

\begin{figure}
\epsscale{.47}
\plotone{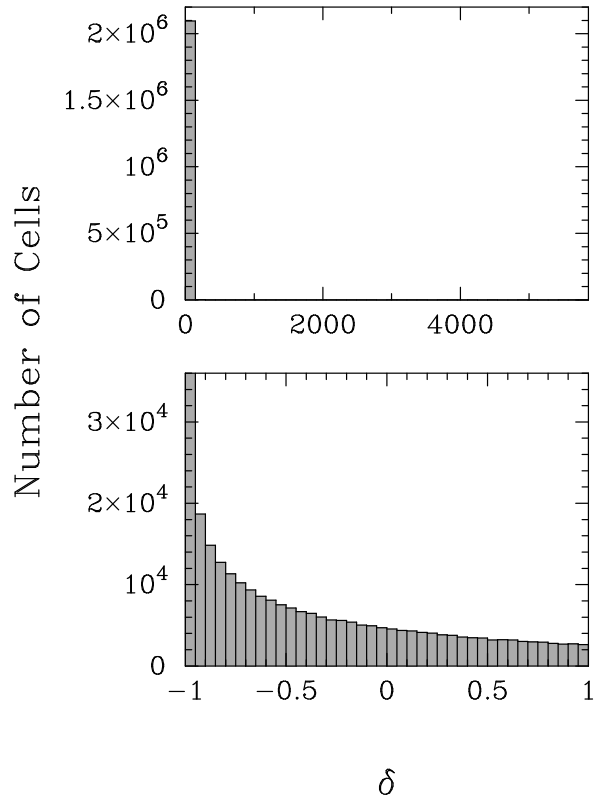}
\caption{Histogram of the density field at redshift $z=0$ for a
         representative simulation.  The histogram is shown for the
         range of values spanned by $\delta$ (top), and for the
         subrange $-1\leq\delta\leq1$ (bottom).  In the bottom panel,
         the upper limit of the $y$-axis is set to 1/50 the maximum
         height of the histogram.}
\end{figure}

\begin{figure}
\epsscale{.47}
\plotone{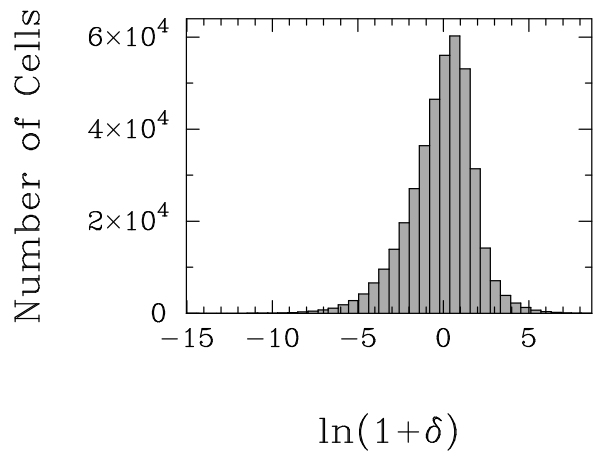}
\caption{Histogram of the transformed density field at redshift $z=0$
         for the same representative simulation as in Fig.\ 4.}
\end{figure}

\begin{figure}
\epsscale{.33}
\plotone{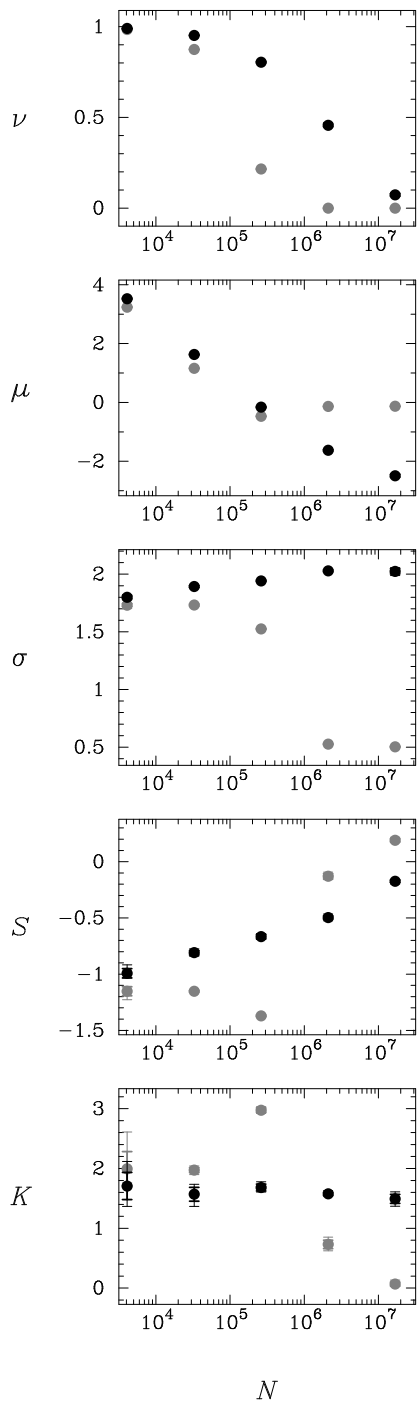}
\caption{Statistics of the transformed density field as functions of
         $N$: the fraction of void cells $\nu$, mean $\mu$, standard
         deviation $\sigma$, skewness $S$ and kurtosis $K$.  The
         redshifts shown are $z=0$ (black) and $z=15$ (grey).  The
         scatter of the diagnostics is shown by plotting their average
         (circles), root-mean-square deviation (thick error bars) and
         range (thin error bars).}
\end{figure}

\begin{figure}
\epsscale{.66}
\plotone{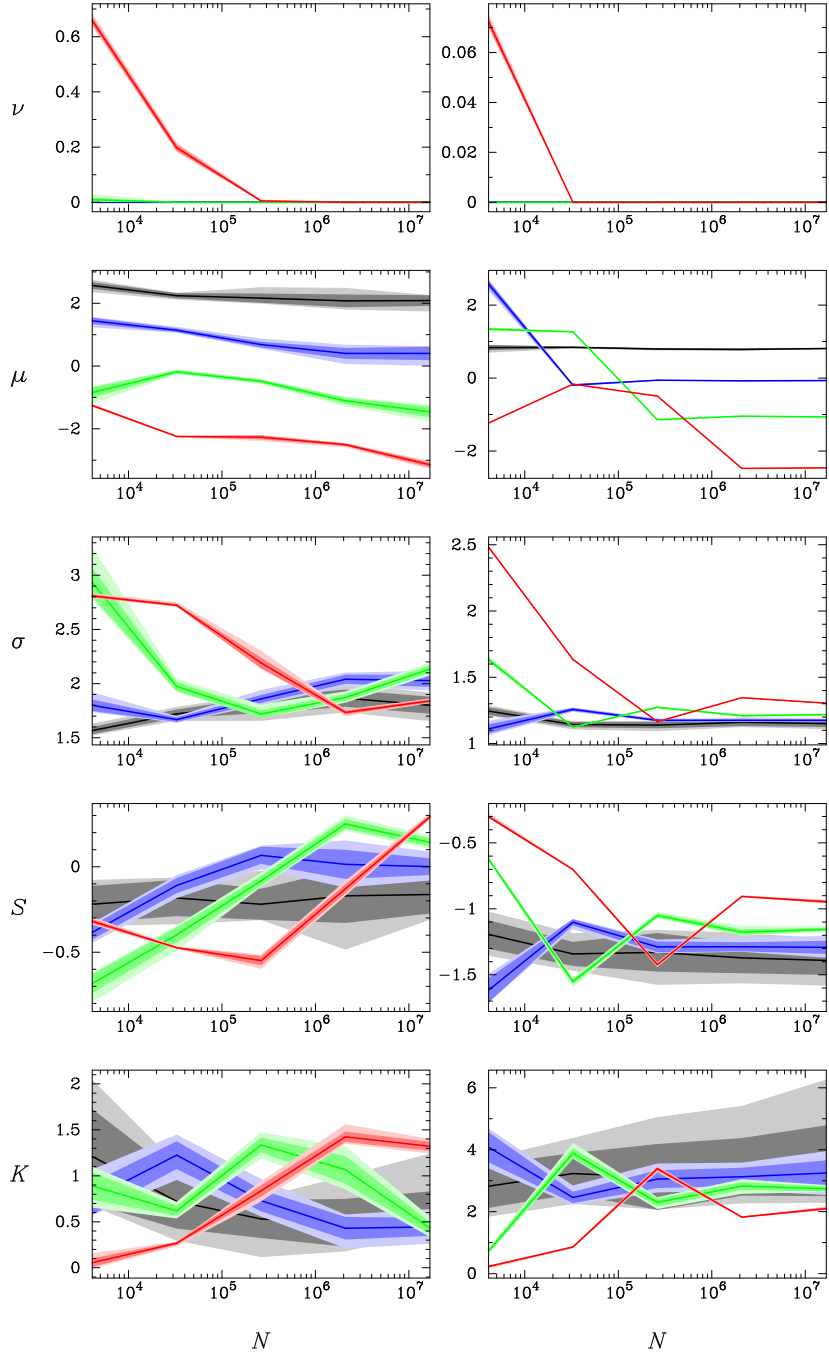}
\caption{Wavelet statistics of the density field as functions of $N$
         at redshifts $z=0$ (left) and $z=15$ (right): the void
         probability $\nu$, mean $\mu$, standard deviation $\sigma$,
         skewness $S$ and kurtosis $K$.  The spatial scales shown are
         $s=1$ (red), $s=2$ (green), $s=3$ (blue) and $s=4$ (grey).
         The scatter of the diagnostics is shown by plotting their
         average (lines), root-mean-square deviation (dark regions)
         and range (light regions).}
\end{figure}

\begin{figure}
\epsscale{.72}
\plotone{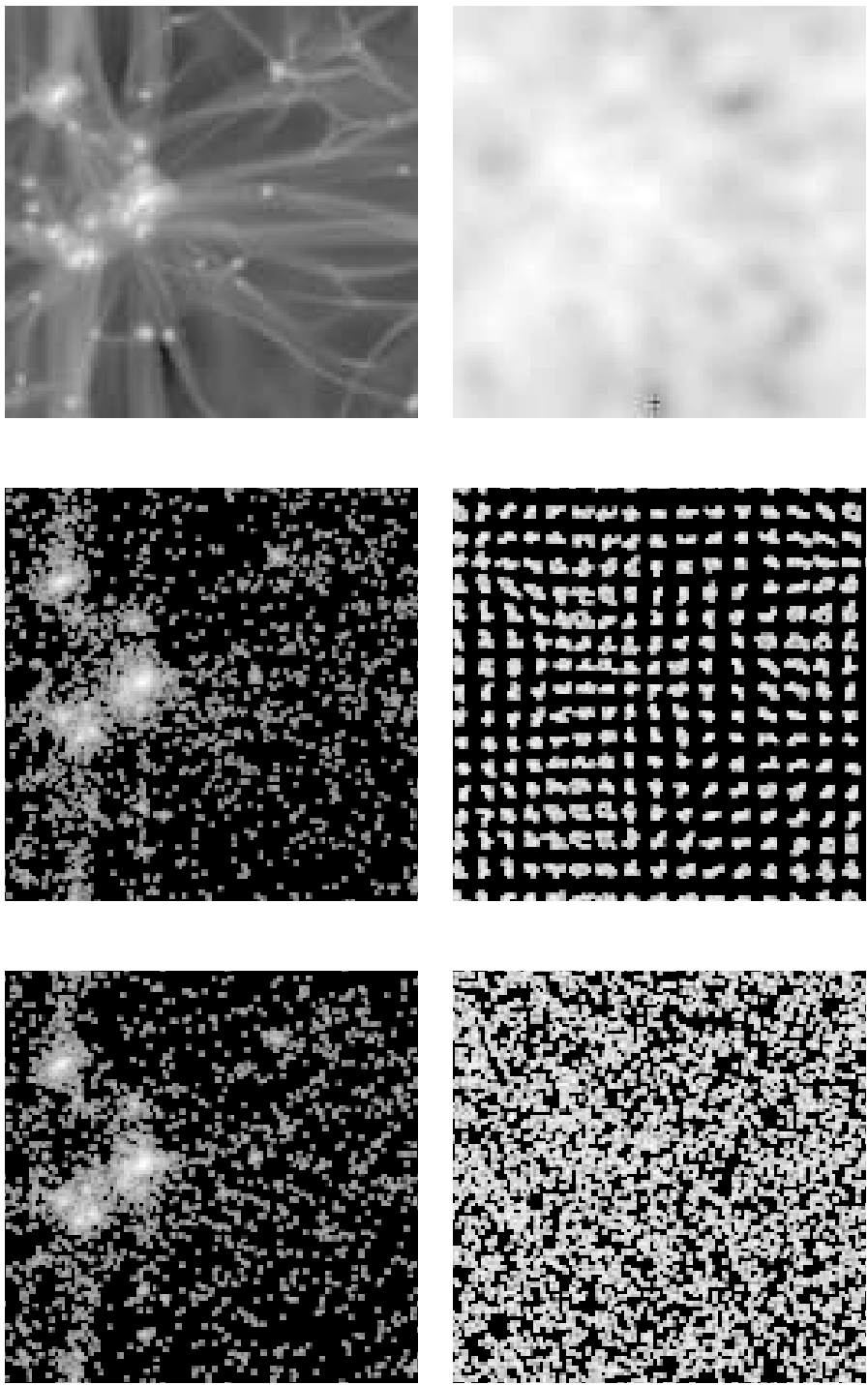}
\caption{Logarithmic grey-scale map of the projected density field at
         redshifts $z=0$ (left) and $z=15$ (right) for the additional
         simulations with power-spectrum truncation (top), regular
         sub-sampling (middle) and random sub-sampling (bottom).}
\end{figure}

\begin{figure}
\epsscale{.6}
\plotone{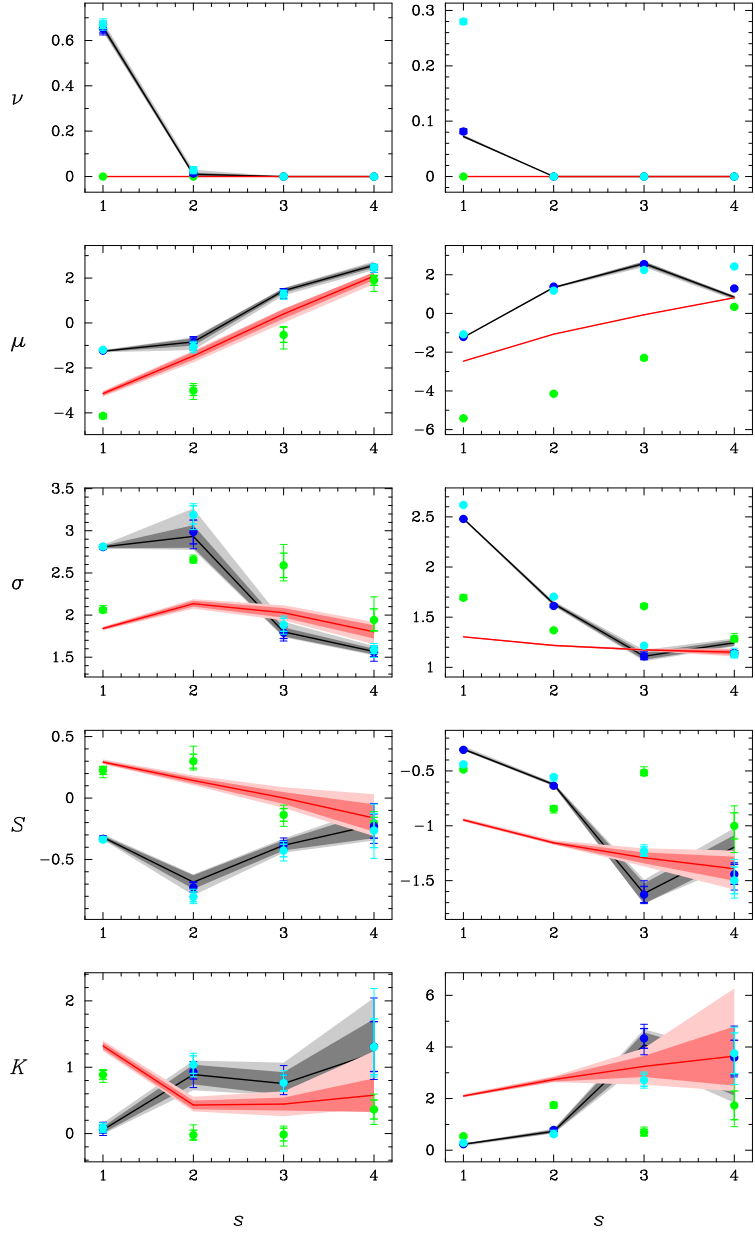}
\caption{Wavelet statistics of the density field as functions of the
         spatial scale $s$ at redshifts $z=0$ (left) and $z=15$
         (right): the void probability $\nu$, mean $\mu$, standard
         deviation $\sigma$, skewness $S$ and kurtosis $K$.  The sets
         of simulations shown are the original simulations with
         $N=16^{3}$ (grey curves) and $N=256^{3}$ (red curves), and
         the additional simulations with power-spectrum truncation
         (green symbols), regular sub-sampling (blue symbols) and
         random sub-sampling (cyan symbols).  The scatter of the
         diagnostics for the original/additional simulations is shown
         by plotting their average (lines/circles), root-mean-square
         deviation (dark regions / thick error bars) and range (light
         regions / thin error bars).}
\end{figure}

\begin{figure}
\epsscale{1.}
\plotone{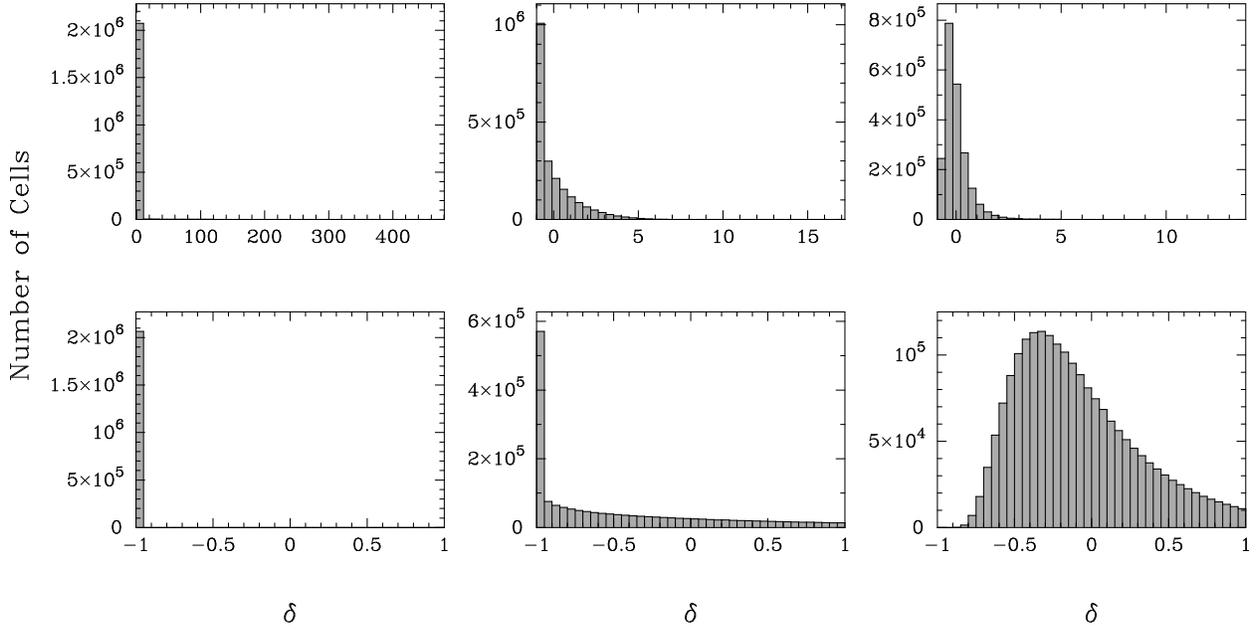}
\caption{Histogram of the initial density field at redshift $z=15$ for
         $N=16^{3}$ (left), $N=64^{3}$ (middle) and $N=256^{3}$
         (right).  Each histogram is shown for the range of values
         spanned by $\delta$ (top), and for the subrange
         $-1\leq\delta\leq1$ (bottom).}
\end{figure}

\begin{figure}
\epsscale{1.}
\plotone{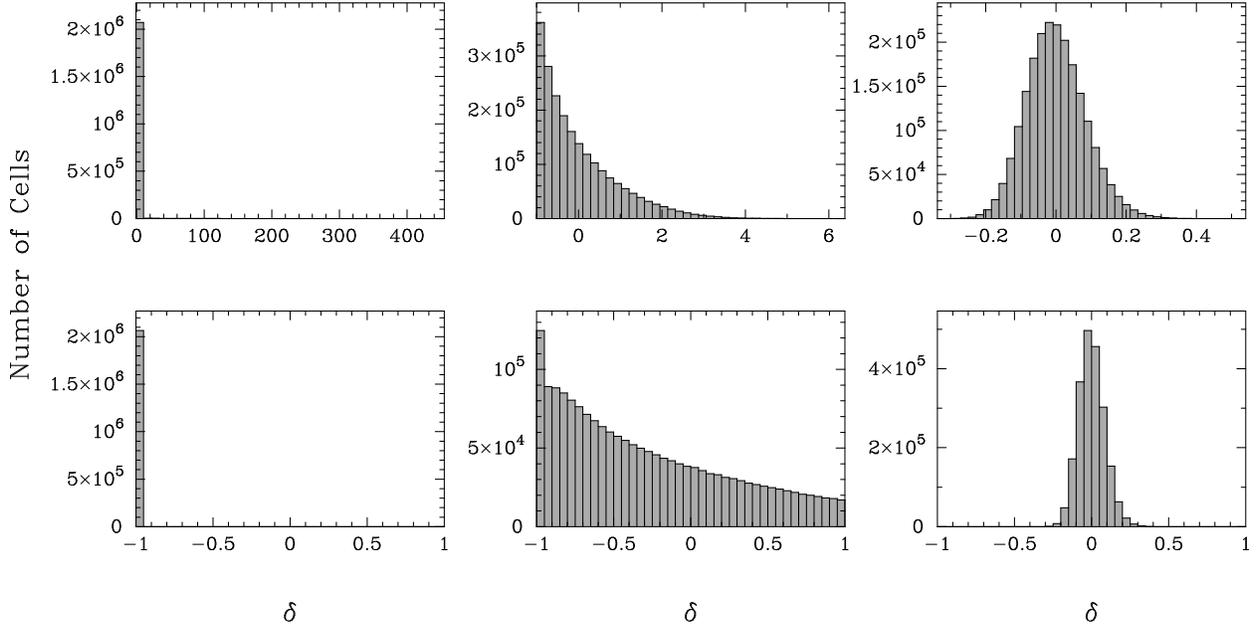}
\caption{Same as Fig.\ 10, but at redshift $z=100$.}
\end{figure}


\begin{thebibliography}{}
\bibitem{} Agertz, O. 2004,
           M\,Sc thesis (Chalmers University of Technology)
\bibitem{} Baertschiger, T., Joyce, M., Sylos Labini, F., \& Marcos,
           B. 2008,
           Phys. Rev. E, 77, 051114
\bibitem{} Bagla, J. S., \& Prasad, J. 2008,
           preprint (arXiv:0802.2796)
\bibitem{} Bate, M. R., \& Burkert, A. 1997,
           MNRAS, 288, 1060
\bibitem{} Binney, J. 2004,
           MNRAS, 350, 939
\bibitem{} Binney, J., \& Knebe, A. 2002,
           MNRAS, 333, 378
\bibitem{} Coles, P., \& Jones, B. 1991,
           MNRAS, 248, 1
\bibitem{} Coles, P., \& Lucchin, F. 2002,
           Cosmology: The Origin and Evolution of Cosmic Structure
           (Chichester: Wiley)
\bibitem{} Col\'{\i}n, P., Valenzuela, O., \& Avila-Reese, V. 2008,
           ApJ, 673, 203
\bibitem{} Dalal, N., Dor\'{e}, O., Huterer, D., \& Shirokov, A. 2007,
           preprint (arXiv:0710.4560)
\bibitem{} Dehnen, W. 2001,
           MNRAS, 324, 273
\bibitem{} Diemand, J., Moore, B., Stadel, J., \& Kazantzidis, S.
           2004,
           MNRAS, 348, 977
\bibitem{} Fang, L.-Z., \& Feng, L.-L. 2000,
           ApJ, 539, 5
\bibitem{} Fang, L.-Z., \& Thews, R. L. 1998,
           Wavelets in Physics (Singapore: World Scientific)
\bibitem{} Feng, L.-L. 2007,
           ApJ, 658, 25
\bibitem{} Grossi, M., Branchini, E., Dolag, K., Matarrese, S., \&
           Moscardini, L. 2008,
           preprint (arXiv:0805.0276)
\bibitem{} Hamana, T., Yoshida, N., \& Suto, Y. 2002,
           ApJ, 568, 455
\bibitem{} Hansen, S. H., Agertz, O., Joyce, M., Stadel, J., Moore,
           B., \& Potter, D. 2007,
           ApJ, 656, 631
\bibitem{} He, P., Feng, L.-L., \& Fang, L.-Z. 2005,
           ApJ, 628, 14
\bibitem{} Heitmann, K., Ricker, P. M., Warren, M. S., \& Habib, S.
           2005,
           ApJS, 160, 28
\bibitem{} Hikage, C., Coles, P., Grossi, M., Moscardini, L., Dolag,
           K., Branchini, E., \& Matarrese, S. 2008,
           MNRAS, 385, 1613
\bibitem{} Hockney, R. W., \& Eastwood, J. W. 1988,
           Computer Simulation Using Particles (Bristol: Institute of
           Physics Publishing)
\bibitem{} Joyce, M., \& Marcos, B. 2007a,
           Phys. Rev. D, 75, 063516
\bibitem{} Joyce, M., \& Marcos, B. 2007b,
           Phys. Rev. D, 76, 103505
\bibitem{} Joyce, M., Marcos, B., \& Baertschiger, T. 2008,
           preprint (arXiv:0805.1357)
\bibitem{} Kayo, I., Taruya, A., \& Suto, Y. 2001,
           ApJ, 561, 22
\bibitem{} Klypin, A., \& Holtzman, J. 1997,
           arXiv:astro-ph/9712217
\bibitem{} Knebe, A., \& Dom\'{\i}nguez, A. 2003,
           Publ. Astron. Soc. Australia, 20, 173
\bibitem{} Knebe, A., Green, A., \& Binney, J. 2001,
           MNRAS, 325, 845
\bibitem{} Knebe, A., Kravtsov, A. V., Gottl\"{o}ber, S., \& Klypin,
           A. A. 2000,
           MNRAS, 317, 630
\bibitem{} Knuth, D. E. 1998,
           The Art of Computer Programming -- Vol. 2: Seminumerical
           Algorithms (Boston: Addison-Wesley)
\bibitem{} Kravtsov, A. V., Klypin, A. A., \& Khokhlov, A. M. 1997,
           ApJS, 111, 73
\bibitem{} Mart\'{\i}nez, V. J., Starck, J.-L., Saar, E., Donoho, D.
           L., Reynolds, S. C., de la Cruz, P., \& Paredes, S. 2005,
           ApJ, 634, 744
\bibitem{} Melott, A. L. 2007,
           preprint (arXiv:0709.0745)
\bibitem{} Moore, B., Quinn, T., Governato, F., Stadel, J., \& Lake,
           G. 1999,
           MNRAS, 310, 1147
\bibitem{} Nelson, A. F. 2006,
           MNRAS, 373, 1039
\bibitem{} Peacock, J. A. 1999,
           Cosmological Physics (Cambridge: Cambridge University
           Press)
\bibitem{} Peebles, P. J. E. 1980,
           The Large-Scale Structure of the Universe (Princeton:
           Princeton University Press)
\bibitem{} Power, C., Navarro, J. F., Jenkins, A., Frenk, C. S.,
           White, S. D. M., Springel, V., Stadel, J., \& Quinn, T.
           2003,
           MNRAS, 338, 14
\bibitem{} Press, W. H., Teukolsky, S. A., Vetterling, W. T., \&
           Flannery, B. P. 2007,
           Numerical Recipes: The Art of Scientific Computing, Third
           Edition (Cambridge: Cambridge University Press)
\bibitem{} Price, D. J., \& Monaghan, J. J. 2007,
           MNRAS, 374, 1347
\bibitem{} Romeo, A. B., Horellou, C., \& Bergh, J. 2003,
           MNRAS, 342, 337 (Paper I)
\bibitem{} Romeo, A. B., Horellou, C., \& Bergh, J. 2004,
           MNRAS, 354, 1208 (Paper II)
\bibitem{} Saar, E., Mart\'{\i}nez, V. J., Starck, J.-L., \& Donoho,
           D. L. 2007,
           MNRAS, 374, 1030
\bibitem{} Sirko, E. 2005,
           ApJ, 634, 728
\bibitem{} Smith, R. E., et al. 2003,
           MNRAS, 341, 1311
\bibitem{} Splinter, R. J., Melott, A. L., Shandarin, S. F., \& Suto,
           Y. 1998,
           ApJ, 497, 38
\bibitem{} Springel, V. 2005,
           MNRAS, 364, 1105
\bibitem{} Stadel, J. G. 2001,
           PhD thesis (University of Washington)
\bibitem{} Stuart, A., \& Ord, J. K. 1991,
           Kendall's Advanced Theory of Statistics -- Vol. 2:
           Classical Inference and Relationship (London: Hodder \&
           Stoughton -- Arnold)
\bibitem{} Teodoro, L., \& Warren, M. S. 2004,
           arXiv:astro-ph/0406174
\bibitem{} Teyssier, R. 2002,
           A\&A, 385, 337
\bibitem{} Tinker, J., Kravtsov, A. V., Klypin, A., Abazajian, K.,
           Warren, M., Yepes, G., Gottl\"{o}ber, S., \& Holz, D. E.
           2008,
           preprint (arXiv:0803.2706)
\bibitem{} Vidakovic, B. 1999,
           Statistical Modeling by Wavelets (New York: Wiley)
\bibitem{} Vogelsberger, M., White, S. D. M., Helmi, A., \& Springel,
           V. 2008,
           MNRAS, 385, 236
\bibitem{} Wang, J., \& White, S. D. M. 2007,
           MNRAS, 380, 93
\bibitem{} Warren, M. S., Abazajian, K., Holz, D. E., \& Teodoro, L.
           2006,
           ApJ, 646, 881
\bibitem{} Wetzstein, M., Nelson, A. F., Naab, T., \& Burkert, A.
           2008,
           preprint (arXiv:0802.4245)
\bibitem{} Zhan, H. 2006,
           ApJ, 639, 617
\end{thebibliography}
\end{document}